\documentclass{iaus}
\usepackage{graphicx}

\newcommand{\params}{\mathcal{P}}
\newcommand{\efold}{$\chi^2$-EF}

\title[Sines, steps and droplets] %% give here short title %%
{Sines, steps and droplets:\\ Semiparametric Bayesian modeling\\
  of arrival time series}

\author[Thomas J. Loredo]   %% give here short author list %%
{Thomas J. Loredo}

\affiliation{Department of Astronomy, Cornell University, \\ 
Ithaca, New York, USA \\
email: {\tt loredo@astro.cornell.edu}}

\pubyear{2012}
\volume{285}  %% insert here IAU Symposium No.
\pagerange{xxx--yyy}
\jname{New horizons in time domain astronomy}
\editors{E. Griffin \& R. Hanisch, eds.}

\begin{document}

\maketitle

\begin{abstract}
I describe ongoing work developing Bayesian methods for flexible modeling of
arrival time series data without binning, aiming to improve detection and
measurement of X-ray and gamma-ray pulsars, and of pulses in gamma-ray
bursts.  The methods use parametric and semiparametric Poisson point
process models for the event rate, and by design have close connections
to conventional frequentist methods currently used in time-domain astronomy.
\end{abstract}

Measuring the arrival times, directions, and energies of individual
quanta---photons or particles---potentially provides the finest possible
resolution of dynamical astronomical phenomena, particularly for high-energy
sources producing low detectable fluxes.  The simplest methods for signal
detection and measurement bin the data for statistical or computational
convenience (e.g., to allow use of asymptotic Gaussian approximations or to
enable fast Fourier decomposition with an FFT).  But methods that instead
directly analyze the event data without binning can detect weaker signals
and probe shorter time scales than methods that require binning.

The ongoing work I briefly describe here is motivated by studies of X-ray
and gamma-ray pulsars, producing periodic signals, and gamma-ray bursts,
producing chaotic signals typically comprised of multiple overlapping
pulses.  In the former case there may be less than one event per period
(particularly in energy-resolved studies); in the latter, time scales as
short as milliseconds are relevant, and detected photons are sparse at high
energies.  Both phenomena motivate development of data analysis
techniques that can milk every hard-won event for what it is worth.

\begin{figure}[t]
% \vspace*{-2.0 cm}
\begin{center}
\includegraphics[width=.8\textwidth]{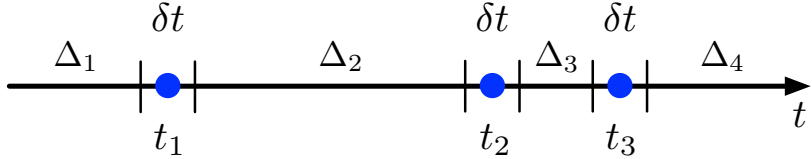} 
% \vspace*{-1.0 cm}
\caption{Arrival time series depicted as points on a timeline, with
detection and nondetection intervals noted.}
\label{timeline}
\end{center}
\end{figure}

For simplicity we here focus just on the arrival time data (also known as
time-tagged event (TTE) data), presuming the events being analyzed have been
selected to have directions consistent with an origin from a single source,
and that energy dependence of any putative signal is not significant (so the
signal's temporal signature is not corrupted by ignoring event energies). We
can represent the data as points on a timeline, as in Fig.~\ref{timeline}.
The dots denote events at times $t_i$ detected within small time intervals,
$\delta t$, representing the instrumental time resolution.  The empty
intervals, denoted $\Delta t_j$, are informative; seeing no events in an
observed interval provides a constraint on the signal, in contrast to simply
not observing during the interval.

Conventional approaches to detecting signals in such data (binned or
unbinned) adopt a frequentist hypothesis testing approach:  one devises a
test statistic that measures departure of the data from the predictions of
an uninteresting ``null'' model, and uses it to see if the null model may be
safely rejected (implying an interesting signal is present).  No explicit
signal model is needed to define such a test.

Here I describe methods developed using the Bayesian approach, where one
compares the null model to explicit alternative models
describing interesting signals.  One of my motivations is to show how
conventional ``alternative free'' test statistics arise in this framework
(exactly or approximately).  This illuminates implicit assumptions
underlying conventional methods; more constructively, it provides a
framework where generalizations of the implicit models may lead to new
methods.  There are several other virtues of adopting a Bayesian approach to
these problems, both pragmatic and conceptual; I briefly note a few below. A
more extensive but still introductory discussion is in Loredo (2011).

We will model the data with a nonhomogeneous Poisson point process in time.
A model, $M$, specifies an intensity function (event rate) $r(t;\params)$
that depends on the model's parameters, $\params$.  Parametric models have a
parameter space of fixed dimension; e.g., a periodic signal model will
typically have frequency, amplitude, and phase parameters, and possibly
additional parameters describing the light curve shape.  Nonparametric
models have a parameter space whose (effective) dimension may grow with
sample size, adapting to the data; it may formally be infinite-dimensional. 
Semiparametric models have a parameter space with a fixed-dimension part
(e.g, the frequency and phase of a periodic model) and a nonparametric part
(e.g., an adaptive light curve shape).

The data drive Bayesian inferences via the likelihood function, the
probability for the data, $D= (\{t_i\}, \{\Delta t_j\})$, given values for
the parameters. Referring to Fig.~\ref{timeline}, we build the likelihood
function by calculating the product of Poisson counting probabilities for
zero counts in the empty intervals, and one count in each detection
interval.  The zero-count probabilities are of the form
$\exp[-\int_{\Delta_i} dt\,r(t)]$, and the one-count probabilities are of the
form $[r(t_i)\delta t] \exp[-r(t)\delta t]$ (presuming $\delta t$ is small
so that $r(t)\delta t\ll 1$).  Thus the likelihood function is
\begin{equation}
\mathcal{L}_M(\params) \equiv p(D|\params,M)
 = \exp\left[-\int_T dt\,r(t)]\right]
   \prod_{i=1}^N r(t_i)\delta t,
\label{like}
\end{equation}
where $T$ denotes the full observing interval and $N$ is the number of
detected events.  To go further, we must specify specific rate models and
priors for the model parameters.  To fit a particular model, we use Bayes's
theorem to calculate a posterior probability for the parameters,
$p(\params|D,M) = p(\params|M) \mathcal{L}_M(\params)/Z_M$, where $Z_M$ is a
normalization constant, given by the integral of the product of the prior
probability density, $p(\params|M)$, and the likelihood function.  To detect
a signal, we instead use Bayes's theorem on a hypothesis space including one
or more models for interesting signals, and the null model (here, a constant
rate model with a single parameter, the amplitude, $A$, with $r(t) = A$). 
In this space, the normalization constant for a particular model, $Z_M$,
plays the role of the likelihood for the model (as a whole); $Z_M$ is often
called the marginal likelihood for the model, where ``marginal'' refers to
the integration over $\params$ used to calculate its value.

As a simple starting point, consider a model where the logarithm of the rate
is proportional to a sinusoid plus a constant; the logarithm guarantees that
the rate itself is non-negative.  Then we may write the rate as
$r(t) = A \exp[\kappa \cos(\omega t - \phi)]/I_0(\kappa)$, where $A$ is the
time-averaged rate, and $I_0(\cdot)$ denotes the modified Bessel function of
order 0 (this normalizes the exponential factor so that $A$ is the
time-average). Light curves with this shape have a single peak per period;
its width (equivalently, the duty cycle) is determined by the concentration
parameter, $\kappa$, with large values corresponding to sharp peaks, and
$\kappa=0$ corresponding to a constant rate.  To estimate the frequency and
concentration, we calculate the posterior for all four parameters,
$p(A,\omega,\kappa,\phi|D,M)$, and marginalize (integrate) over $A$ and
$\phi$.  Adopting a flat prior for the phase, and nearly any prior for $A$
that is independent of the other parameters, we find a marginal posterior
probability density for frequency and concentration proportional to
$I_0[\kappa R(\omega)]/[I_0(\kappa)]^N$,
where $R(\omega)$ is the Rayleigh statistic, given by
\begin{equation}
R^2(\omega) 
  = {1 \over N}\left[\left(\sum_{i=1}^N \cos \omega t_i \right)^2 +
      \left(\sum_{i=1}^N \sin \omega t_i \right)^2\right].
\label{rayleigh}
\end{equation}
Estimation of $\omega$ alone, accounting for uncertainty in all other
parameters, is found by further integrating over $\kappa$;
this is easy to do numerically.  Detection requires calculating the
marginal likelihood, corresponding to a further integration over
$\omega$.  This must be done numerically.  It can be time-consuming for
blind searches, but not significantly more so than the kind of frequency
grid searching done with conventional tests.

The Rayleigh statistic was invented for the well-known (frequentist)
Rayleigh test for periodic signals in arrival time series (see Lewis 1994
for a review of the Rayleigh test and other frequentist
period detection methods mentioned below). The quantity $2R^2(\omega)$ is
called the Rayleigh power; it is the point process analog of the periodogram
or Fourier power spectral density.  From a Bayesian point of view,
the Rayleigh test implicitly assumes periodic signals may be well-modeled
by log-sinusoid rate functions.  Notably, there is no parameter corresponding
to $\kappa$ in the Rayleigh test; also, in practice it is known to
work well only for smooth light curves with a single broad peak per
period.  Such light curves correspond to values of $\kappa$ near
unity, another implicit assumption of the Rayleigh test.  These results
indicate that one can implement Bayesian period searches using
conventional computational tools already at hand for the Rayleigh test.
They also indicate that explicit consideration of the $\kappa$ parameter
may lead to procedures more sensitive to sharply-peaked light curves
than the conventional Rayleigh.

% by analogy to a Fourier series...

Many pulsars have light curves with two or more peaks per period.  This
suggests generalizing the log-sinusoid model to a log-Fourier model,
with the logarithm of the rate proportional to a sum of harmonic
sinusoids.  Adopting a finite sum of $m$ harmonics with concentration
parameters $\kappa_k$ ($k=1$ to $m$, with the fundamental corresponding to
$k=1$), we may proceed analogously to the above analysis.  The larger number
of phases and concentration parameters thwarts analytical integration; in an
approximate treatment the posterior distribution for frequency and
concentration is proportional to $\exp[S(\omega)]$ with
$S(\omega) \equiv \sum_{k=1}^m \kappa_k R(k\omega)$.  The frequentist test
generalizing the Rayleigh test to multiple harmonics is the $Z^2_m$ test,
with $Z^2_m(\omega) = 2\sum_{k=1}^m R^2(k\omega)$, a sum of Rayleigh powers at
harmonics.  Notably, the Bayesian analysis uses the sum of $R(k\omega)$
values (``harmonic Rayleigh amplitudes'') rather than powers.  Note that,
for $\kappa_k = 1$,
$S^2(\omega) = Z^2_m(\omega)/2 + \sum_k\sum_{j\ne k} R(j\omega)R(k\omega)$;
that is, $S^2$ contains information not in $Z^2_m$.  Roughly speaking,
$Z^2_m$ corresponds to incoherently summing power in harmonics, but the
quantity arising in the Bayesian treatment of a harmonic model instead sums
amplitudes, accounting for phase information ignored by $Z^2_m$.

A popular frequentist period detection method that aims to be sensitive
to periodic signals of complex shape is $\chi^2$ epoch folding (\efold).
One folds the arrival times modulo a trial period to produce a phase,
$\theta_i$, for each event, with $\theta_i$ in the interval $[0,2\pi]$.
For a constant signal, the phases should be uniformly distributed.  The
\efold\ method bins the phases into $B$ bins and uses Pearson's $\chi^2$
to test consistency of the binned phases with a uniform distribution.
This motivates a Bayesian model with a piecewise-constant rate function
with $B$ steps per period.  Gregory \& Loredo (1992; GL92)
analyzed this model, with the rate in a bin given by $Af_k$, where $A$
is the average rate, and the shape parameters, $f_k$ ($k=1$ to $B$),
specify the fraction of the rate attributed to each bin, with
$\sum_k f_k = 1$.  GL92 assigned a constant prior to the shape parameters,
based on the intuition that this spreads probability across all possible
shapes.  After marginalizing over the shape parameters, the posterior
for frequency and phase is inversely proportional to the multiplicity
of the set of counts of events in phase bins.  In a large-count limit, this
is approximately $\exp(-\chi^2/2)$, providing a tie to \efold.  The method
performed impressively, detecting an X-ray pulsar where the Rayleigh test
failed, and performing well in a simulation study by Rots (1993) comparing
it to other methods.

Despite these successes, there is room for improvement in the
GL92 analysis, for a surprising reason.  The constant prior
adopted in GL92 in fact does not spread probability over all possible
shapes.  As the number of bins increases, the constant prior puts ever
larger probability in the neighborhood of flat models, making it harder than
necessary to detect narrow peaks.  The reason is a ``curse of
dimensionality'' known as {\em concentration of measure}:  a
multidimensional distribution built out of the product of broad
one-dimensional distributions with finite moments concentrates its
probability in a decreasing volume of parameter space as dimension
increases.  Concentration can be avoided by letting the parameters of the
one-dimensional component distribution vary with the target dimension.
A theoretically appealing way to do this is to require {\em divisibility}
of the prior, e.g., the four-bin prior should reduce to the two-bin
prior if we create a two-bin model out of the combination of bins
1 and 2, and 3 and 4.  A divisible Dirichlet distribution prior accomplishes
this, and improves sensitivity to sharply peaked light curves
so long as any constant background component is small (Loredo 2011).

Extending the construction to functions described with an infinite number of
bins or points leads one to consider {\em infinitely divisible} priors
for nonparametric functions: Gaussian process priors for curve fitting,
Dirichlet process priors for modeling probability densities, and L\'evy
process priors for modeling Poisson intensities.  With a team of
statistician and astronomer colleagues, I am developing methods using priors
built on L\'evy processes for modeling pulses in gamma-ray bursts.  This
approach can quantify uncertainty even in the regime where pulses are highly
overlapping.  Its implementation involves compound Poisson processes, as
arise in simple models of accumulation of rain, where drops with a
distribution of sizes fall radomly over a region of space.  And so Bayesian
modeling of arrival time series has led us from sines to steps to droplets.

Some of the research summarized here is currently supported by grant
NNX09AK60G from NASA's Applied Information Systems Research Program.  I am
grateful to my collaborators on this grant, Mary Beth Broadbent, Carlo
Graziani, Jon Hakkila, and Robert Wolpert, for their ongoing contributions
to this research.

\end{document}